\documentclass[letter]{aa} % for the letters 
\usepackage{graphicx,url}
\usepackage[normalem]{ulem}
\usepackage{mathrsfs,amssymb,amsmath}
\usepackage[varg]{txfonts}
\usepackage{booktabs}
\usepackage{multirow}
\usepackage[breaklinks, colorlinks, citecolor=blue]{hyperref}

\begin{document}

   \title{The Orion-Taurus ridge: a synchrotron radio loop at the edge of the Orion-Eridanus superbubble}
    
\titlerunning{A synchrotron radio loop at the edge of the Orion-Eridanus superbubble}
   \authorrunning{Bracco et al.}

   \author{Andrea Bracco \inst{1}, Marco Padovani\inst{2}, and Juan D. Soler\inst{3}
     }
   \institute{Laboratoire de Physique de l'Ecole Normale Sup\'erieure, ENS, Universit\'e PSL, CNRS, Sorbonne Universit\'e, Universit\'e de Paris, F-75005 Paris, France \\
   \email{andrea.bracco@phys.ens.fr} 
   \and
   INAF--Osservatorio Astrofisico di Arcetri, Largo E. Fermi 5, 50125 Firenze, Italy 
   \and
   Istituto di Astrofisica e Planetologia Spaziali (IAPS), INAF, Via Fosso del Cavaliere 100, 00133 Roma, Italy 
       }

   \date{Received June 26, 2023; accepted August 31, 2023}

% \abstract{}{}{}{}{} 
% 5 {} token are mandatory
 
  \abstract
{
Large-scale synchrotron loops are recognized as the main source of diffuse radio-continuum emission in the Galaxy at intermediate and high Galactic latitudes. Their origin, however, remains rather unexplained.

Using a combination of multi-frequency data in the radio band of total and polarized intensities, for the first time in this letter, we associate one arc -- hereafter, the Orion-Taurus ridge -- with the wall of the most prominent stellar-feedback blown shell in the Solar neighborhood, namely the Orion-Eridanus superbubble. 

We traced the Orion-Taurus ridge using 3D maps of interstellar dust extinction and column-density maps of molecular gas, $N_{\rm H_2}$. We found the Orion-Taurus ridge at a distance of 400\,pc, with a plane-of-the-sky extent of $180$\,pc. Its median $N_{\rm H_2}$ value is  $(1.4^{+2.6}_{-0.6})\times 10^{21}$ cm$^{-2}$. 

Thanks to the broadband observations below 100 MHz of the Long Wavelength Array, we also computed the low-frequency spectral-index map of synchrotron emissivity, $\beta$, in the Orion-Taurus ridge. We found a flat distribution of $\beta$ with a median value of $-2.24^{+0.03}_{-0.02}$ that we interpreted in terms of depletion of low-energy ($<$ GeV) cosmic-ray electrons in recent supernova remnants ($10^5$ - $10^6$ yrs). Our results are consistent with plane-of-the-sky magnetic-field strengths in the Orion-Taurus ridge larger than a few tens of $\mu$G ($> 30 - 40 \,\mu$G).   

We report the first detection of diffuse synchrotron emission from cold-neutral, partly molecular, gas in the surroundings of the Orion-Eridanus superbubble.
This observation opens a new perspective to study the multiphase and magnetized interstellar medium with the advent of future high-sensitivity radio facilities, such as the C-Band All-Sky Survey and the Square Kilometre Array.

}

               
   \maketitle
%
%-------------------------------------------------------------------

\section{Introduction}\label{sec:intro}

The intermediate-to-high Galactic latitude sky brightness in radio continuum is dominated by numerous large-scale loops that have long been observed \citep[e.g.,][]{berkhuijsen1971a,Haslam1971} and interpreted as synchrotron emission structures \citep{vidal15,planckXXV2016,Thomson2021,Martire2023}, but whose origin remains uncertain. Most of them have been associated with remnants of supernova (SN) explosions \citep{Berkhuijsen1971b,Salter1983}, where cosmic-ray electrons (CRe) are accelerated in shock-compressed magnetic fields in the interstellar medium (ISM), giving rise to synchrotron emission \citep{Ferriere1991a}. The superposition of such loops from SN remnants may also be responsible for most of the total radio emission along the Galactic plane itself, as postulated by \citet{Mertsch2013}. However, to date, a reference model on the physical origin of synchrotron radio loops is still lacking \citep[e.g.,][]{Sofue2015,Kataoka2021,Panopoulou2021}. 

In this letter, we present the emission and the physical properties of the Orion-Taurus ridge, which is a 25-degree-wide radio loop located in the constellation of Orion and centered around Galactic coordinates $(l,b)=(190^{\circ},-20^{\circ})$. Despite being already detected \citep[e.g.,][]{Milogradov1997, Borka2008, vidal15, planckXXV2016}, the Orion-Taurus ridge represents a rather unreported radio structure compared to more renown loops, such as Loop I, II, III, and IV \citep{berkhuijsen1971a,vidal15}. 

\begin{figure*}[!h]
\begin{center}
\resizebox{1\hsize}{!}{\includegraphics{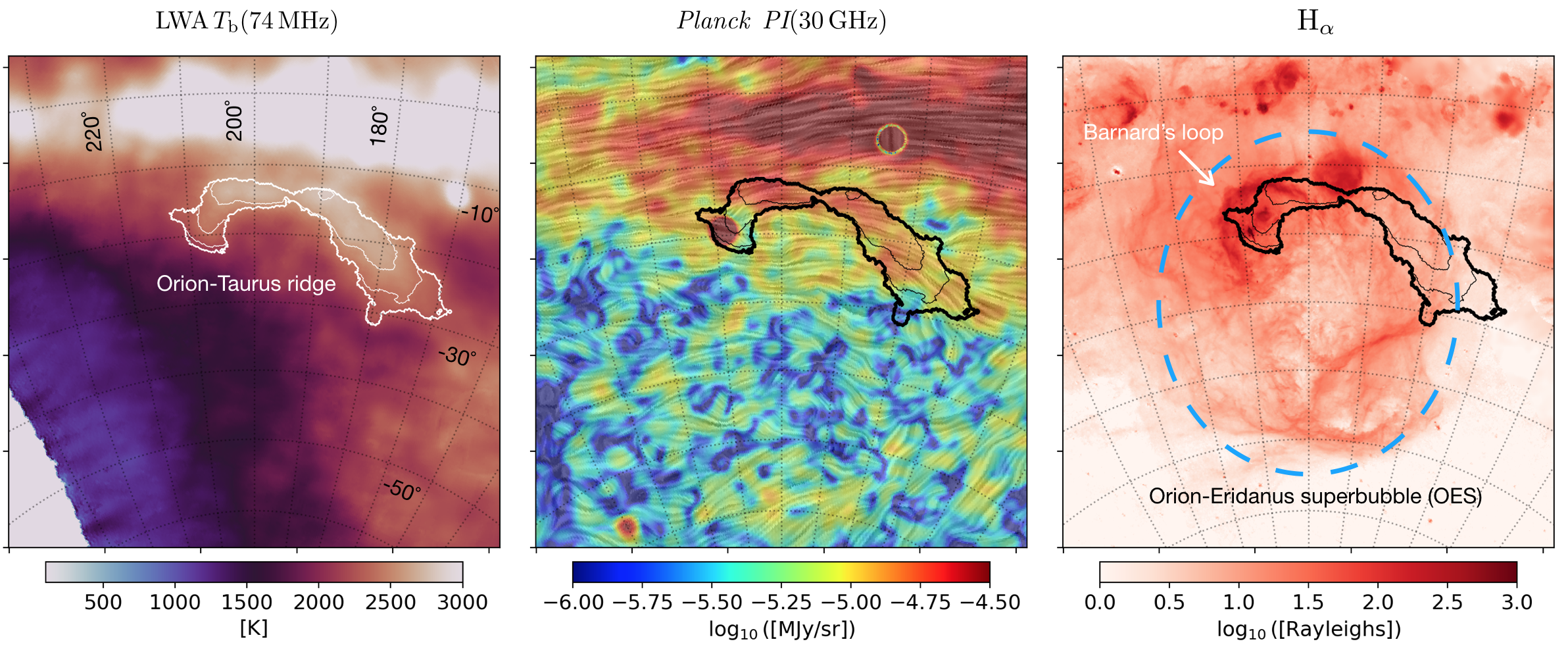}}
\caption{ Stereographic projections showing the Orion-Taurus ridge. {\it Left:} Brightness temperature ($T_{\rm b}$) at 74\,MHz observed by the LWA telescope at 2$^{\circ}$ FWHM resolution. The contours are at 2100, 2200, 2400, 2600, and 2800\,K, respectively. {\it Center:} Polarized intensity ($PI$) at 30\,GHz observed by the {\it Planck} satellite at the same FWHM resolution of $T_{\rm b}$(74\,MHz) with the plane-of-the-sky magnetic field orientation traced by line integral convolution \citep{Cabral1993}. Black contours are the same as in the left panel. {\it Right:} H$\alpha$ emission map from \citet{Finkbeiner2003} showing the Orion-Taurus ridge in contours and the extent of the Orion-Eridanus superbubble (OES) traced in blue as in \citet{Soler2018}. The same Galactic coordinate grid is shown in all panels with steps in $l$ and $b$ of $10^{\circ}$ and centered in ($l,b$) = ($200^{\circ},-30^{\circ}$). The stereographic projection was produced with the Python routine {\tt stereo}$\_${\tt proj}$\_${\tt hpx} that can be found at  \protect\url{http://github.com/abracco/cosmicodes/blob/master/4GMIMS/Planck_routines.py}.} 
\label{fig:Arc}
\end{center}
\end{figure*}

For the first time, we show that the Orion-Taurus ridge is a synchrotron loop associated with cold neutral ISM at the northern edge of the Orion-Eridanus superbubble \citep[OES,][]{Heiles1976,Reynolds1979,Ochsendorf2015}.  
The OES is one of the closest-by (distance, $D<500$ pc) and largest circular structures in the sky (see the H$\alpha$ map from \citet{Finkbeiner2003} in the right panel of Fig.~\ref{fig:Arc}). It includes the Orion molecular cloud (OMC) and is considered the joint result of strong ionizing UV radiation, stellar winds, and SN explosions from the Orion OB1 association \citep[e.g.,][]{Heiles1997,Bally2008}. 

In the following, we introduce the multi-frequency data used in the analysis (Sect.~\ref{sec:data}). We detail the morphological correlation between radio data of total and polarized intensities tracing synchrotron in the Orion-Taurus ridge and the dusty, partly molecular, ISM at the edge of the OES (Sect.~\ref{sec:methods}). We compute the spectral index of synchrotron emissivity, $\beta$, in the Orion-Taurus ridge using radio continuum data below 100\,MHz (Sect.~\ref{sec:spec}). Based on the parametric model of \citet{Padovani2018} for the energy spectrum of CRe responsible for the low-frequency radio emission \citep{Padovani2021}, we propose an original mechanism to explain the values of $\beta$ in the Orion-Taurus ridge. 

\section{Data}\label{sec:data}
In this section, we describe the multi-wavelength data set used to probe the Orion-Taurus ridge.

\subsection{Synchrotron total emission: Long Wavelength Array data}\label{ssec:lwa}

We used data from the northern-sky survey of the Long Wavelength Array (LWA)\footnote{\url{http://lda10g.alliance.unm.edu/LWA1LowFrequencySkySurvey/}} as described in \citet[][hereafter D17]{Dowell2017}. The original brightness temperature,$T_{\rm b} (\nu)$, maps and their uncertainties correspond to nine frequency bands between 35 and 80\,MHz with angular resolution between FWHM $\approx5^{\circ}$ and $\approx2^{\circ}$, respectively (see Table~1 in D17). These maps trace diffuse synchrotron emission in the Galaxy.

In this work, we only focused on the five highest frequency bands, namely at 50, 60, 70, 74, and 80\,MHz, in order to reach FWHM of $w_0 = 3.5^{\circ}$, at worst. 
All LWA maps, in units of K, are downloadable in {\tt HEALPix}\footnote{\url{http://healpix.sf.net}} format \citep{Gorski2005} with $N_{\textrm side}$\,$=$\,256 (13.7$\arcmin$ wide pixels). In Sect.~\ref{sec:spec}, where we study the synchrotron spectral index of the loop, we smoothed the maps to the common resolution of $w_0$, assuming Gaussian beams\footnote{The effective FWHM is $\sqrt{w_0^2-w_i^2}$, where $w_i$ represents the Gaussian FWHM of each original map.} and using the {\tt healpy.sphtfunc.smoothing} Python package \citep{Zonca2019}. 

The Orion-Taurus ridge can be seen in the left panel of Fig.~\ref{fig:Arc}, where, as an example, we show it with white contours on top of the $T_{\rm b} (74~{\rm MHz})$ map. The Orion-Taurus ridge was isolated by unsharp masking the $T_{\rm b}$(74 MHz) map using a Gaussian window of FWHM\,$=$\,$10^{\circ}$ \citep[see also][]{vidal15}, and masking out all filtered values below 40\,K. The choice of this threshold was only used to optimize the sky selection of the Orion-Taurus ridge that was finalized with the function {\tt label} of the {\tt scipy} Python package. The Orion-Taurus ridge extends over several tens of degrees both in Galactic longitude and latitude.   
 
\subsection{Synchrotron polarization: {\it Planck} data}\label{ssec:planck}

Because of the numerous instrumental and physical challenges that affect low-frequency radio-polarimetric data, such as those of LWA \citep{Taylor2012} -- e.g., Stokes-parameter leakage, ionospheric and interstellar Faraday rotation -- it was easier to spot the Orion-Taurus ridge in synchrotron polarization at high frequency from satellite observations. We inspected the 30 GHz {\it Planck} maps of Stokes $Q$ and $U$ \citep{planckI2020}\footnote{\protect\url{http://pla.esac.esa.int/}\label{planckdata}} for the Orion-Taurus ridge, after smoothing them at FWHM\,$=$\, $2^{\circ}$. In the right panel of Fig.~\ref{fig:Arc}, we show the Orion-Taurus ridge with black contours on the corresponding polarized intensity ($PI$\,$=$\,$ \sqrt{Q^2+U^2}$). It was already discovered by \citet{planckXXV2016} but never discussed in depth. Although $PI$ extends more toward the east, the visual correlation with most of the loop seen in $T_{\rm b}$(74\,MHz) supports its origin as a synchrotron structure. Like most of the radio loops \citep{vidal15}, the Orion-Taurus ridge also appears elongated along the plane-of-the-sky (POS) magnetic field, $\vec{B}_{\perp}$, which we inferred from synchrotron polarization rotating by 90$^{\circ}$ the 30-GHz polarization angle, $\psi$\,$=$\,$0.5 \arctan(-U,Q)$ \footnote{The minus sign converts the {\it Planck} Stokes U to the IAU convention.}. The $\vec{B}_{\perp}$ orientation is shown using the line integral convolution \citep[LIC,][]{Cabral1993} from {\tt healpy}.

In this work, we considered {\it Planck} data to trace the total gas column density ($N_{\rm H}$) along the line of sight (LOS). In particular, we used the $N_{\rm H}$ map derived from the dust optical depth at 353 GHz ($\tau_{353}$), which was obtained by fitting a modified blackbody spectrum to the {\it Planck} intensity data below 857 GHz and IRAS at 100 $\mu$m \citep{planckXI2014}. The $\tau_{353}$ map (FWHM = 5$\arcmin$) was converted into $N_{\rm H}$ using a conversion factor of $8.3\times 10^{25}$ cm$^{-2}$ \citep{PIPXXXV2016}. The $N_{\rm H}$ map is shown in a composite image in Fig.~\ref{fig:Arc_ext}. 

Finally, as discussed in Sect.~\ref{sec:spec}, we used the {\it Planck} emission measure ($EM$) map to estimate the contribution of free-free absorption to synchrotron emission. The maximum-likelihood (ML) value of the $EM$ map$^{\rm \ref{planckdata}}$ was the result of a Bayesian analysis for component separation \citep{planck2016X}. We smoothed the original map to the resolution of $w_0$.

\subsection{Orion-Taurus ridge distance: 3D dust map}\label{ssec:3dust} 

We made use of the publicly available 3D dust map from \citet{Lallement19}\footnote{\url{http://cdsarc.u-strasbg.fr/viz-bin/qcat?J/A+A/625/A135}} to constrain the distance to the Orion-Taurus ridge. This data cube was obtained by combining \emph{Gaia} DR2 and 2MASS photometric data with \emph{Gaia} DR2 parallaxes. The LOS dust distribution was inferred with a regularized Bayesian tomographic inversion of starlight extinction measurements. This technique has the advantage of accounting for spatial correlation of dust structures with a limited physical resolution of 25\,pc within 1\,kpc from the Sun. 
For our purpose, however, we did not require high resolution in the 3D dust reconstruction \citep[e.g.,][]{Green19,Leike20}. The 25\,pc spacing corresponds to the value of $w_0$ at a distance of 400\,pc, which is approximately the distance of the OMC \citep[e.g.,][]{Grosschedl2021}. The 3D dust map contains values of differential extinction (${\rm d}A_{\rm V}/{\rm d}D$) at 5500\,{\AA} in units of ${\rm mag~pc^{-1}}$. Examples of the integrated 3D dust maps are shown in Fig.~\ref{fig:Arc_ext} and Fig.~\ref{fig:ext_maps}.

\section{Morphological correlation analysis}\label{sec:methods}

\begin{figure}[!h]
\begin{center}
\resizebox{1.0\hsize}{!}{\includegraphics{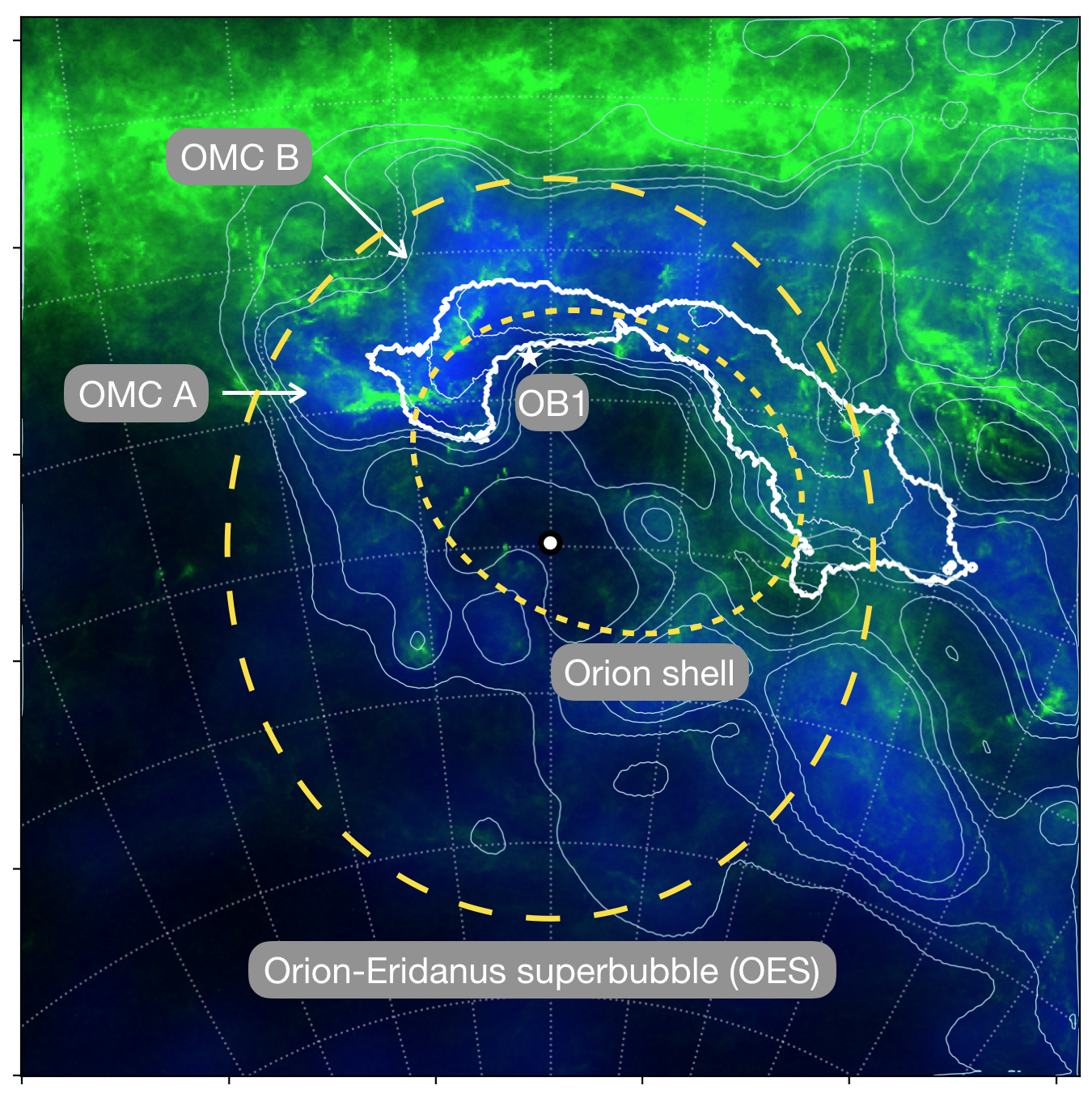}}
\caption{ Stereographic view of the 3D dust map integrated between 300 and 500\,pc (in blue) and the total gas column density ($N_{\rm H}$, in green) toward the Orion-Taurus ridge, shown with white contours. Light-blue contours trace the 3D dust map at $A_{\rm V} = 2, 3, 4, 5$ mag. The $N_{\rm H}$ map is shown above $10^{21}$ cm$^{-2}$. 
The coordinate grid and the white contours are the same as in Fig.~\ref{fig:Arc}. 
The long-dashed yellow line indicates the edge of the OES.
The short-dashed-yellow line indicates the edge of the Orion shell reported by \citet{Foley2023}. A white circle highlights the center of the OES.}
\label{fig:Arc_ext}
\end{center}
\end{figure}
The Orion-Taurus ridge is seen both in $T_{\rm b}$(74 MHz) and in $PI$(30 GHz) (see Fig.~\ref{fig:Arc}). It is a synchrotron structure with curvature directed toward the center of the OES, or possibly of a subset of it at a comparable distance named the Orion shell \citep[see Fig.~\ref{fig:Arc_ext} and][]{Foley2023}. 

In Fig.~\ref{fig:Arc_ext} we show a composite image, where the tentative edge of the superbubble \citep[e.g.,][]{Soler2018} is sketched in yellow on top of both the $N_{\rm H}$ map (in green) and the 3D dust map integrated between 300 and 500\,pc (${A}_{\rm V}(300,500{\rm pc})$, in blue). The Orion-Taurus ridge, delimited by white contours as in Fig.~\ref{fig:Arc}, closely follows the northern dusty edge of the OES. It reveals a coherent morphological correlation with the 3D dust map extending from the OMC (A and B) to the more diffuse western rim with ${A}_{\rm V}(300,500{\rm pc}) \gtrsim 5$ mag and $N_{\rm H} \gtrsim 10^{21}$\,cm$^{-2}$. The mean value of ${A}_{\rm V}(300,500{\rm pc})$ within the Orion-Taurus ridge is 8 mag, which integrated over 200\,pc corresponds to a lower limit\footnote{This is only a lower limit because most likely the gas distribution is not uniform within 200\,pc, but condensed on a smaller length scale.} to the mean number density ($n_{\rm H}$) of $n_{\rm H} \approx 50$\,cm$^{-3}$, typical of the cold neutral medium \citep{Ferriere2020}.   
As shown in Appendix~\ref{app:ext}, this visual correlation would not be attained if the 3D dust map was integrated in different distance ranges, between 100 and 300 pc or between 500 and 700 pc (see Fig.~\ref{fig:ext_maps}).

\begin{figure}[!h]
\begin{center}
\resizebox{1.\hsize}{!}{\includegraphics{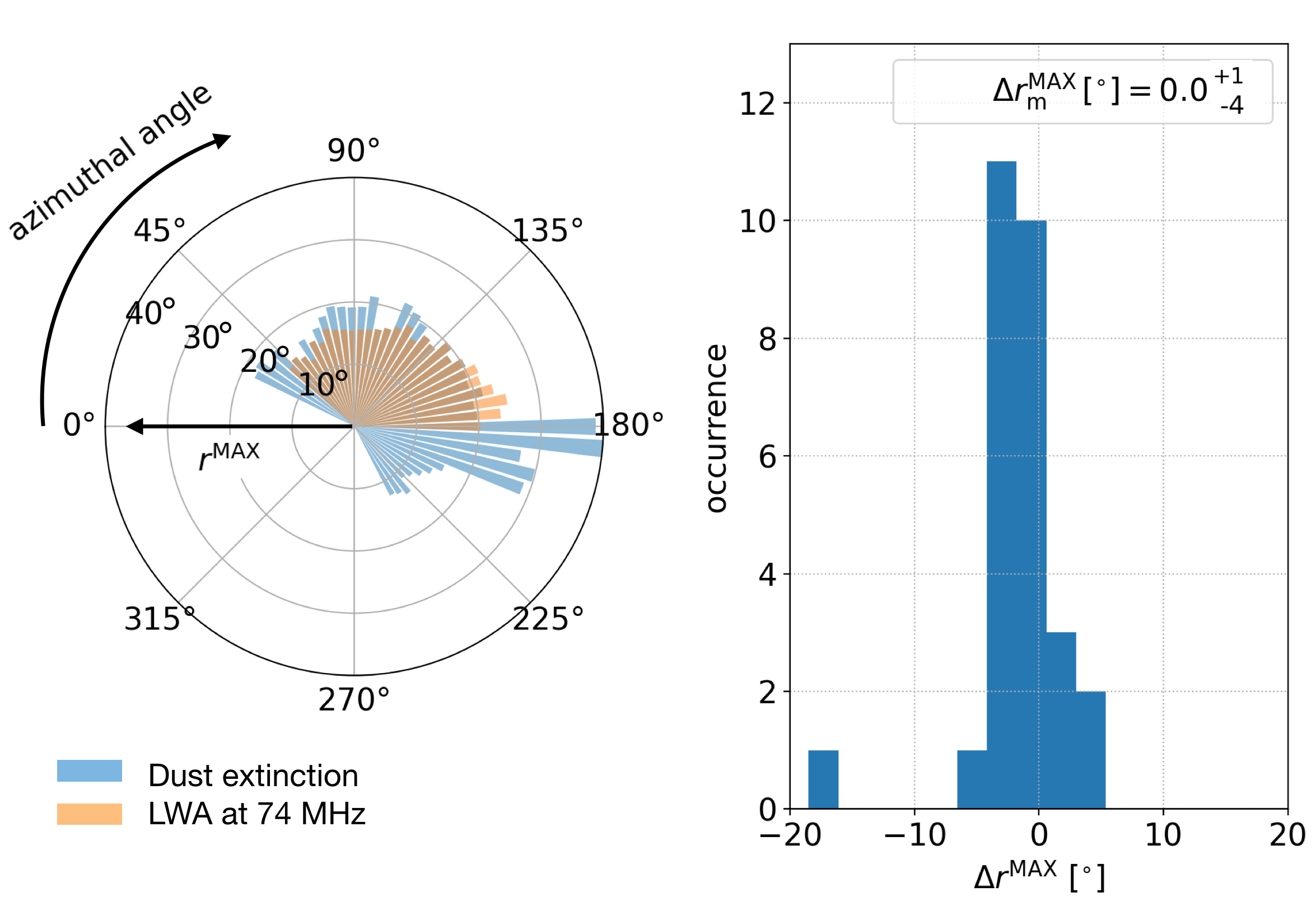}}
\caption{Morphological 
correlation between the Orion-Taurus ridge and the 3D dust map. {\it Left panel:} Polar plot of $r^{\rm MAX}$ as a function of the azimuthal angle in bins of 5$^{\circ}$ (increasing in the E-W direction). $r^{\rm MAX}$ is computed for $T_{\rm b}$(74 MHz) (orange) and the 3D dust map integrated between 300 and 500 pc (blue). {\it Right panel:}  Histogram of the difference between the $r^{\rm MAX}$ distributions shown on the left.
The reported values correspond to the median and the corresponding errors (16th and 84th percentiles).}
\label{fig:orie}
\end{center}
\end{figure}

We quantified the correlation by comparing the radial profiles of the 3D dust map with that of $T_{\rm b}$(74 MHz) from the center of the OES (see the white circle in Fig.~\ref{fig:Arc_ext}). We defined the plane-of-the-sky distance ($r^{\rm MAX}$, in units of degrees) from the center of the superbubble to the first peak of a given radial profile. The peak was found using the {\tt find-peaks} function of {\tt scipy}. Note that $r^{\rm MAX}$ was only considered if the peak of the radial profiles was at least 20\% of the maximum value of the corresponding map. In Fig.~\ref{fig:orie}, we show a polar plot encoding the distributions of $r^{\rm MAX}$ as a function of the azimuthal angle increasing from $0^{\circ}$ to $360^{\circ}$ following the E-W direction. We binned the azimuthal angle in steps of $5^{\circ}$. The plot compares $r^{\rm MAX}$ defined at 74 MHz (orange) with that of the 3D dust map (blue). The Orion-Taurus ridge only correlates, over $135^{\circ}$ between $45^{\circ}$ and $180^{\circ}$, with the structure of the 3D dust map toward the northern rim of the OES at about 400\,pc (see Appendix~\ref{app:ext} for comparison with other distance ranges). For this particular case, in the right panel of Fig.~\ref{fig:orie}, we also show the difference between the two $r^{\rm MAX}$ distributions, which peaks at $0^{\circ}$ with a dispersion smaller than $5^{\circ}$. The median value, the 16th, and 84th percentile errors are $\Delta r^{\rm MAX}_{\rm m} = 0.0^{+1}_{-4}$ $^{\circ}$. 

Our morphological analysis established the correlation between the Orion-Taurus ridge, seen in total and polarized synchrotron intensities, with cold neutral medium traced by dust extinction. We also checked the molecular gas column density map, $N_{{\rm H}_2}$, inferred from observations of atomic hydrogen, H{\small{I}}, emission spectra presented in \citet{Kalberla2020}. The Orion-Taurus ridge shows up at $N_{{\rm H}_2} \gtrsim 10^{21}$\,cm$^{-2}$ with a median value of $ (1.4^{+2.6}_{-0.6})$\,$\times$\,$10^{21}$\,cm$^{-2}$. To our knowledge, this is the first detection of diffuse synchrotron emission from molecular gas along the northern edge of the OES.   

\section{Synchrotron spectral index map}\label{sec:spec}
\begin{figure}[!h]
\begin{center}
\resizebox{1.0\hsize}{!}{\includegraphics{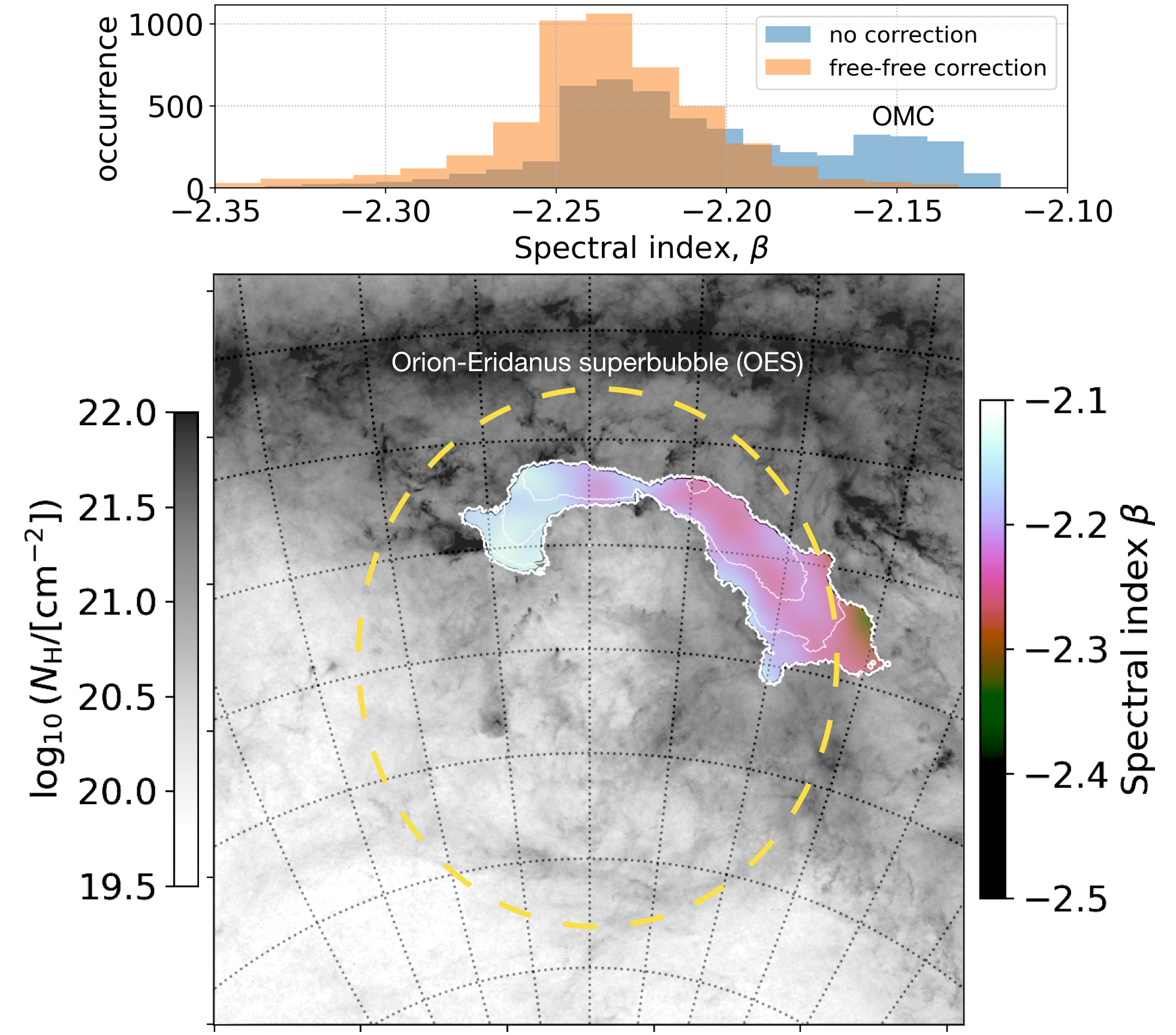}}
\caption{Synchrotron spectral-index map ($\beta$, FWHM $\approx 3.5^{\circ}$, colors) of the Orion-Taurus ridge computed using LWA data at 50, 60, 70, 74, and 80 MHz. The $\beta$ map is overlaid on the $N_{\rm H}$ map from {\it Planck}. The inset above shows the histograms of $\beta$ before (blue) and after (orange) free-free absorption correction. The OMC corresponds to the blue peak at $\beta = -2.15$.}
\label{fig:beta}
\end{center}
\end{figure}

Thanks to the broadband frequency coverage of the LWA, we could study in detail the synchrotron properties of the Orion-Taurus ridge. We used low-frequency data as synchrotron represents the only relevant mechanism in this radio range, without the free-free contamination in emission typical of the high-frequency end ($>$\,1\,GHz). 
Given the five LWA bands introduced in Sect.~\ref{ssec:lwa}, we computed the spectral index of synchrotron emissivity, $\beta$\,$= $\,${\rm d} \ln (T_{\rm b}(\nu))/{\rm d} \ln (\nu)$. The map is shown in Fig.~\ref{fig:beta}, where all the displayed values have a signal-to-noise ratio larger than 5. Our $\beta$ map is consistent with that reported in D17 at a lower angular resolution, which was obtained based on all nine LWA bands (see figure~11). These values of $\beta$ are slightly flatter than those found at 5$^{\circ}$ FWHM by \cite{Guzman2011} -- ranging between $-2.35$ and $-2.45$, see their Fig.~6. Despite this difference, as in  \cite{Guzman2011}, we also found a similar gradient between the main body of the arc and the location of the OMC. As shown by the blue histogram on the top of Fig.~\ref{fig:beta}, the OMC peaks at $\beta = -2.15$ with generally flatter values than the rest of the arc, which has a median value of $\beta_{\rm m} = -2.21^{+0.06}_{-0.03}$.  

\subsection{$\beta$-flattening from free-free}\label{ssec:freefree}

One possible source of flattening of the synchrotron spectral index below 100 MHz comes from contamination of bremsstrahlung, free-free, absorption along the LOS \citep{Dowell2017, Stanislavsky2023}. This effect was first hypothesized in studies of radio spectra of SN remnants \citep{Kassim1989} and proposed to infer the properties of the ionized gas in the Galaxy from extended envelopes of \hbox{H\,{\sc ii}} regions to the warm ionized medium \citep[WIM,][]{Reynolds1998}. Because of the scattering of radio photons with ionized gas along the LOS the observed synchrotron emission may be reduced by a factor $e^{-\tau_{\rm ff}(\nu)}$, where $\tau_{\rm ff}(\nu)$ represents the free-free optical depth at frequency $\nu$. Following \citet{Stanislavsky2023}, this can be expressed as
\begin{equation}\label{eq:tauff}
    \tau_{\rm ff}(\nu) = 3.014\times10^4 Z \left( \frac{T_e}{[{\rm K}]} \right) ^{-1.5}\left( \frac{\nu}{[{\rm MHz}]} \right)^{-2} \left(\frac{EM}{[{\rm cm^{-6}\, pc}]}\right) g_{\rm ff},
\end{equation}
where $Z$ is the average number of ion charges, $T_e$ is the electron temperature, and $EM$ is the emission measure, that is, the integral along the LOS of the electron density, $n_e$, squared (i.e., $EM = \int_{\rm LOS} n^2_{e}{\rm d}l$). The $g_{\rm ff}$ value is the Gaunt factor that we assume equal to $\ln{(49.55/Z/\nu)} + 1.5\ln{T_e}$ \citep{Stanislavsky2023}.

Considering the vicinity of the Orion-Taurus ridge with the Orion OB1 association, where most of the hydrogen ($Z=1$) must be photodissociated by strong UV radiation, we applied Eq.~\ref{eq:tauff} with an average $T_e = 10^4$ K \citep{Ochsendorf2015} and the $EM$ map from $\it Planck$ (see Sect.~\ref{sec:data}). After correction of the LWA $T_{\rm b}$ maps, we derived the spectral index distribution shown in orange in Fig.~\ref{fig:beta}, with a median value of $\beta^{\rm c}_{\rm m} = -2.24^{+0.03}_{-0.02}$, where "c" stands for ``corrected''. If the OMC peak is strongly suppressed after the free-free correction, the average $\beta$ distribution remains the same. We propose another mechanism contributing to lowering the observed $\beta$ in LWA data, which relates to variations of the CRe energy spectrum possibly caused by interacting shocks in the Orion-Taurus ridge and variability of emission sources.

\subsection{Proposed mechanism for explaining $\beta$}\label{ssec:beta}

At a given frequency, the synchrotron spectral index depends on both the CRe energy spectrum and the strength of $\vec{B}_{\perp}$. Synchrotron emission below a few hundred MHz is mostly produced by CRe with energies < GeV for $B_{\perp}$ of a few $\mu$G \citep{Padovani2021}. Their exact energy spectrum, however, remains uncertain. Considering local measurements of the CRe energy flux \citep[e.g.,][]{Orlando18}, and typical $B_{\perp}$ in the ISM, it is difficult to account for the flat $\beta$ found in this work \citep[see Fig.~9 in ][where $\beta \le -2.48$ at 60 MHz for $B_{\perp}$ $\approx$ 30 $\mu$G]{Padovani2021}.     

We modelled the low-energy part of the CRe spectrum using the parametric approach presented in \cite{Padovani2018} and fixed the high-energy part (> 10 GeV) to the interstellar value. This choice is supported by high-energy measurements of the CR-proton spectrum in the OES with the Fermi-LAT satellite that found no significant difference with respect to the diffuse ISM \citep{Joubaud2020}.

At low energy, as detailed in Appendix~\ref{app:CRe}, using Eq.~\ref{eq:je} we could depart from standard interstellar conditions \citep{Orlando18} and control the population of CRe. We were only able to reproduce the range of $\beta$ observed with LWA by severely damping the population of low-energy CRe ($\sim$100 MeV to GeV) and considering strong $B_{\perp}$ larger than a few tens of $\mu$G. These values of $B_{\perp}$ are almost one order of magnitude larger than the average LOS magnetic-field strength of $\approx$ $-3\,\mu$G measured with Zeeman splitting of H{\small{I}} lines toward the OMC \citep{Heiles1997}.  

Although these results highly depend on our limited knowledge of the CRe energy spectrum in the Galaxy, the depletion of low-energy CRe can be related to the specific conditions of the Orion-Taurus ridge as part of the OES (see Sect.~\ref{sec:methods}). In the context of CRe properties in SN remnants, recent work by \cite{Morlino2021} showed that a spectral break in the CRe energy spectrum below a few GeV should be expected as a consequence of the drop in acceleration efficiency of the CRe during the adiabatic stage of the remnant expansion. This is a plausible explanation as the OES is considered the result of highly episodic star formation in the last 10-15 Myrs. Several SN explosions contributed to shaping the cavity and driving the global dynamics of the structure \citep[e.g.,][]{Bally2008}. 
Some examples follow. One recent SN remnant ($\sim$$10^5$\,yrs) was identified by \citet{Ochsendorf2015} and considered at the origin of the H$\alpha$-structure known as 
Barnard's loop (see right panel in Fig.~\ref{fig:Arc}). \citet{Foley2023} also considered a newly discovered cavity in the 3D dust map from \citet{Leike20}, called the Orion shell (see Fig.~\ref{fig:Arc_ext}), which was probably generated by a couple of SN explosions in the last few Myrs. 

\begin{figure}[!h]
\begin{center}
\resizebox{1.0\hsize}{!}{\includegraphics{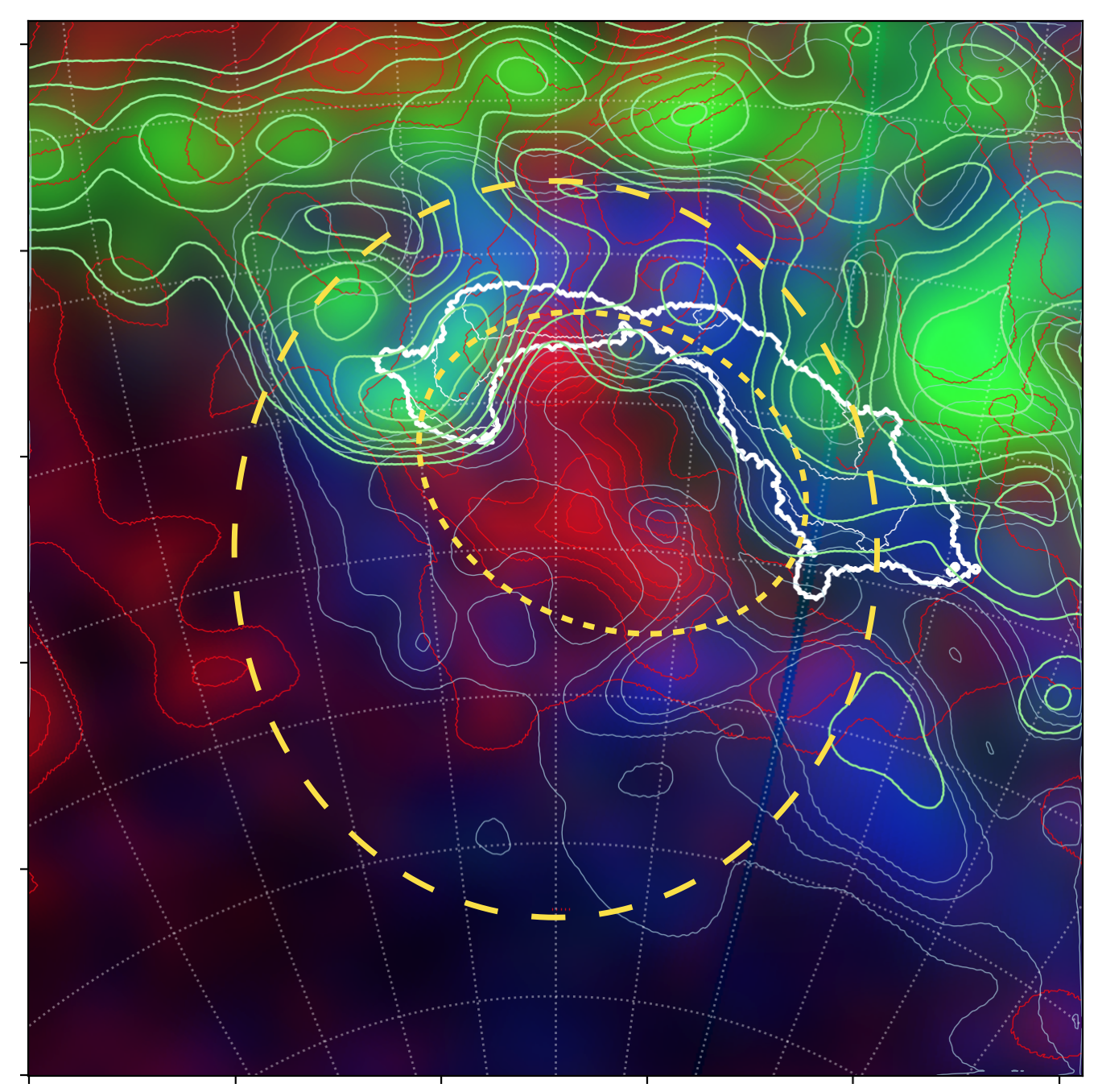}}
\caption{ RGB composite image showing: the 3D dust map integrated between 300 and 500\,pc in blue colours and contours as in Fig.~\ref{fig:Arc_ext}; the $^{26}$Al emission map from \citet{Diehl2002} in red colours and contours. Red contours are between  
$2.8\times10^{-6}$ and $2.4\times10^{-4}$\,cm$^{-2}$\,sr$^{-1}$\,s$^{-1}$ in steps of $2.7\times10^{-5}$\,cm$^{-2}$\,sr$^{-1}$\,s$^{-1}$.
The Orion-Taurus ridge is shown with white contours as in Fig.~\ref{fig:Arc}; the $N_{\rm H_2}$ map from \citet{Kalberla2020} is in green colours and contours between $6.3\times 10^{20}$ and $10^{22}$ cm$^{-2}$ in $\log_{10}$-steps of 0.2. The yellow circles are the same as in Fig.~\ref{fig:Arc_ext}.}
\label{fig:orionshell}
\end{center}
\end{figure}

This hypothesis is supported by the presence of strong $^{26}$Al emission in the region, which has a half-life of 0.7\,Myr mostly traces recent feedback events \citep[e.g.,][]{Foley2023}. In Fig.~\ref{fig:orionshell}, we show the 3D dust map integrated between 300 and 500~pc combined with the $^{26}$Al map based on COMPTEL $\gamma$-ray data at 1.809\,MeV \citep{Diehl2002} also presented in \citet{Foley2023}. The image clearly reveals the $^{26}$Al emission (in red) within the cavity seen in dust extinction, encompassed by both the Orion-Taurus ridge (white contours) and the molecular gas traced by the $N_{\rm H_2}$ map (in green) from \citet{Kalberla2020} smoothed at the resolution of $w_0$.

Compression by multiple SN episodes may be at the origin of the organized structure of $\vec{B}_{\perp}$ in the arc and of its large strength. Similar values of $B_{\perp}$ in the superbubble (of the order of tens of $\mu$G) were also found along its southern Galactic edge \citep{Soler2018,Joubaud2019}. Moreover, as shown in Appendix~\ref{app:CRe}, using a $B_{\perp}$ value of a few tens of $\mu$G ($> 30 - 40\,\mu$G) and CRe energy spectra that account for the $\beta$ distribution in the Orion-Taurus ridge, we were able to recover the same order of magnitude of $T_{\rm b}$(74\,MHz) and $PI$(30~GHz) as measured by LWA and {\it Planck}. Their median values and corresponding errors are $2534^{+166}_{-210}$\,K and $(1.5\pm 1)\times 10^{-5}$\,MJy sr$^{-1}$, respectively (see Figs.~\ref{fig:Arc} and ~\ref{fig:PI}).

\section{Summary and discussion}\label{sec:summary}

In this work, we presented the first detection of diffuse synchrotron emission from cold-neutral, dusty, and partly molecular gas. We combined data from the LWA telescope \citep{Dowell2017}, the {\it Planck} satellite \citep{planckI2020}, and the 3D dust maps of \citet{Lallement19}. 

We discovered the physical correlation between one radio loop -- the Orion-Taurus ridge -- seen both in total intensity below 100 MHz and in polarized intensity at 30 GHz, with the density structure of the northern edge of the OES as traced by interstellar dust extinction. The multi-wavelength correlation extends over several tens of degrees on the sky ($\sim 25^{\circ}$) with dust located between 300 and 500\,pc from the Sun. Considering an average distance of 400\,pc, the Orion-Taurus ridge is about $\sim 180$ pc wide, which is compatible with the expected edge-size of superbubbles from SN feedback \citep[e.g.,][]{Ntormousi2017,Grudic2021}. We note that, in addition to being consistent with the dusty edge of the OES \citep{Ochsendorf2015,Pon2016}, the Orion-Taurus ridge could also be associated with the recently-discovered Orion shell \citep{Foley2023}.

Whether the Orion-Taurus ridge is linked to the OES or to the Orion shell, it is most likely still the result of synchrotron radiation from CRe in relatively dense shock-compressed gas ($N_{\rm H_2} \gtrsim 10^{21}$\,cm$^{-2}$) produced by several episodes of recent SN remnants. 
This scenario is supported by enhanced $^{26}$Al emission seen in the proximity of the Orion-Taurus ridge \citep{Diehl2002, Foley2023} and by our analysis of the spectral index of synchrotron emissivity -- $\beta$ -- between 50 and 80 MHz from the LWA. After correcting for possible contamination of free-free absorption along the sightline, we found a rather flat distribution of $\beta$ in the Orion-Taurus ridge with a median value equal to $-2.24^{+0.03}_{-0.02}$. 
Using the parametric model of \citet{Padovani2018}, we explained such flat $\beta$ values by damping the low-energy part of the CRe spectrum below 1 GeV. This phenomenological fact may be associated with the expected drop in acceleration efficiency of CRe below a few GeV following the adiabatic phase of SN remnants \citep{Morlino2021}. 
Because of strong magnetic-field amplification caused by compression \citep{Ferriere1991b}, SN feedback could also explain the organized structure of the plane-of-the-sky magnetic field (see Fig.~\ref{fig:Arc}) and its strength, $B_{\perp}$, that we derived from the spectral-index analysis of the order of a few tens of $\mu$G ($> 30 - 40 \,\mu$G). These values are consistent with those already estimated in the southern edge of the OES \citep[e.g.,][]{Soler2018} and with the amount of total and polarized synchrotron emissions from the LWA telescope and the {\it Planck} satellite, respectively. 

This work highlights the link between the magnetized and cold ISM by means of low-frequency diffuse synchrotron emission, extending the studies that link radio-interferometric data in polarization below 200\,MHz with tracers of the neutral ISM, such as H{\small{I}} spectroscopic data, interstellar dust extinction, and dust thermal polarized emission \citep{vanEck2017,Jelic2018, Bracco2020b,Turic2021,Bracco2022}. 
However, the case of the Orion-Taurus ridge is the first involving synchrotron total intensity. Possibly, this is due to the vicinity of the OES to the solar neighborhood, to the ongoing high-mass star formation in the region, and to its Galactic location with very low background emission. 
The advent of future high-sensitivity facilities in the radio spectrum, such as the C-Band All-Sky Survey \citep{Jones2018} and the Square Kilometre Array \citep{Dewdney2009}, will likely begin revealing structures such as the Orion-Taurus ridge in a more systematic way. This will open a new window in our understanding of the complex relationship between cosmic rays, magnetic fields, and interstellar gas over a wide range of Galactic environments \citep{Dickinson2015, PadovaniGalli2018}. 

\begin{acknowledgements}
We thank the anonymous referee for the constructive report that improved this manuscript. 
We are grateful to F. Mertens for introducing us to the LWA data. We warmly thank D. Galli, G. Morlino, D. Arzoumanian, I. Grenier, and F. Boulanger for insightful discussions and comments. We appreciated the kindness of M. Foley, J. Alves, and R. Diehl for sharing with us the COMPTEL $^{26}$Al map \citep{Diehl2002}. This work would not have been possible without the inspiring babbling of Elio and the tireless patience of Delphine. 
A.B. acknowledges support from the European Research Council through the Advanced Grant MIST (FP7/2017-2022, No. 742719). Some of the results in this paper have been derived using the {\tt healpy} and {\tt HEALPix} packages. 
JDS acknowledges funding by the European Research Council via the ERC Synergy Grant ``ECOGAL -- Understanding our Galactic ecosystem: From the disk of the Milky Way to the formation sites of stars and planets'' (project ID 855130).
In the analysis we made use of {\tt astropy} \citep{astropy2018}, {\tt scipy} \citep{Virtanen2020}, and {\tt numpy} \citep{Harris2020}.
\end{acknowledgements}

\bibliographystyle{aa}
\bibliography{aanda.bbl}

\appendix
\section{Distinct structures in the 3D dust map}\label{app:ext}
The Orion-Taurus ridge under study only shows morphological correlation with the 3D dust map at the position of the northern edge of the OES ($\approx 400$ pc). 

\begin{figure}[!h]
\begin{center}
\resizebox{1\hsize}{!}{\includegraphics{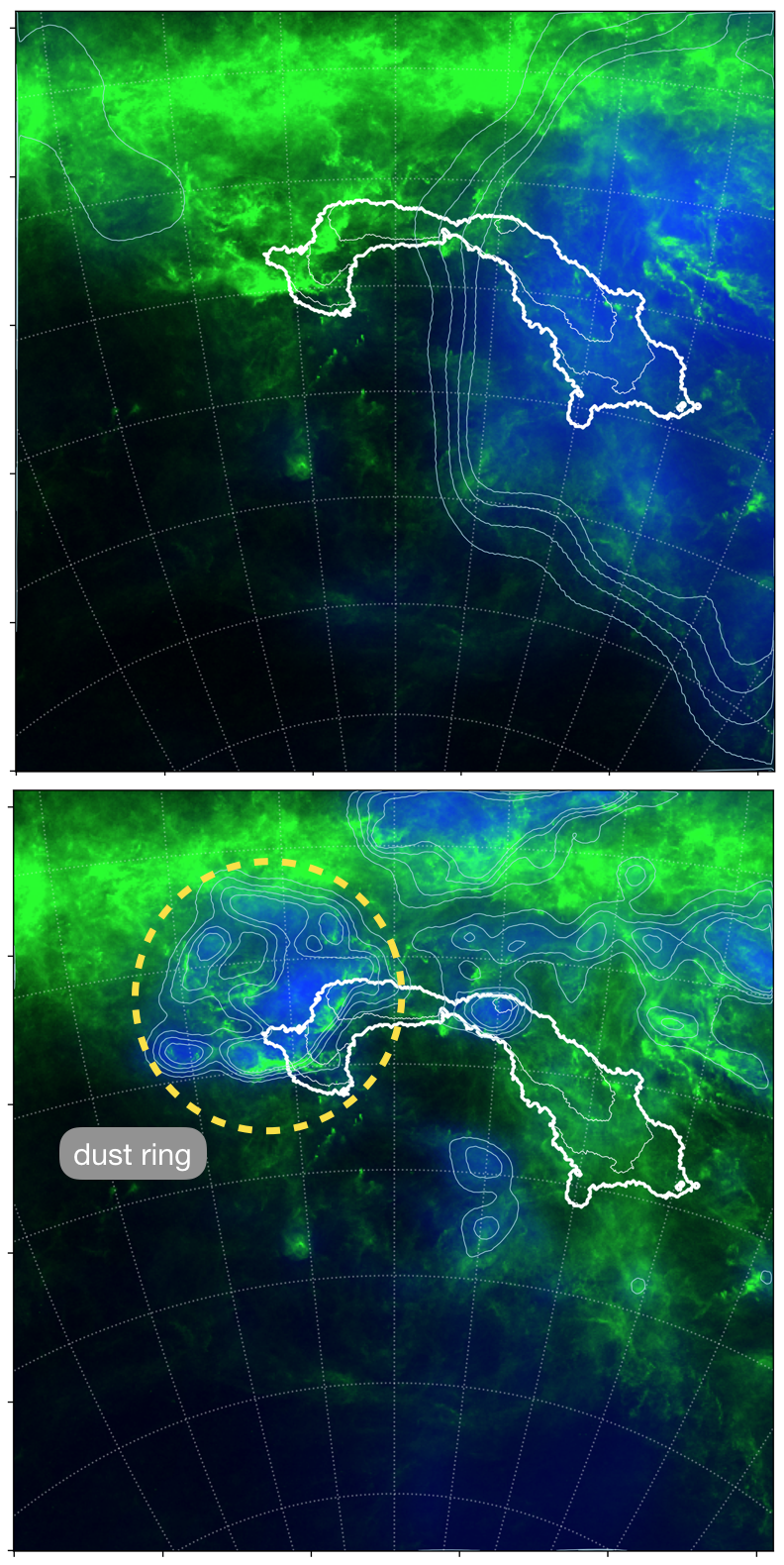}}
\caption{Same as in Fig.~\ref{fig:Arc_ext} but for two different distance ranges, namely between 100 and 300 pc (top panel) and between 500 and 700 pc (bottom panel). 
%In the bottom panel a tenuous shell-like structure, the {\it Orion N shell}, in the North of the OMC can be seen with a yellow dashed line.
}
\label{fig:ext_maps}
\end{center}
\end{figure}

\begin{figure}[!h]
\begin{center}
\resizebox{1.\hsize}{!}{\includegraphics{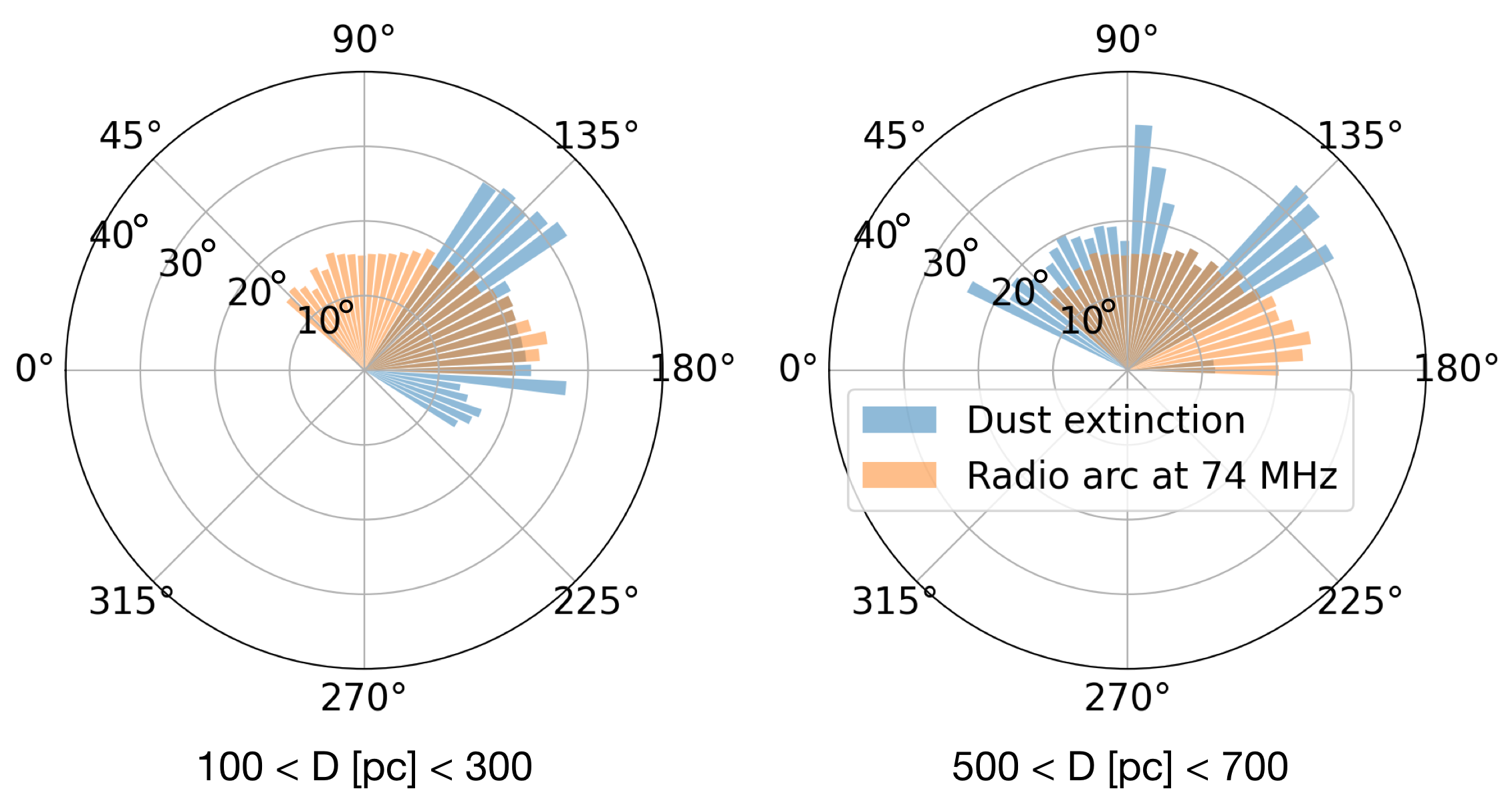}}
\caption{Same as in the left panel of Fig.~\ref{fig:orie} but for two distinct distance ranges.}
\label{fig:no_corr}
\end{center}
\end{figure}

In this section we show the visual comparison (Fig.~\ref{fig:ext_maps}) and the $r^{\rm MAX}$ distributions (Fig.~\ref{fig:no_corr}) between the Orion-Taurus ridge defined at 74 MHz and the 3D dust map integrated at different locations along the LOS: between 100 and 300 pc and between 500 and 700 pc. The structures in the maps are uncorrelated with $\Delta r^{\rm MAX}_{\rm m} = 14^{+2}_{-17}$ $^{\circ}$ and $\Delta r^{\rm MAX}_{\rm m} = -3^{+14}_{-8}$ $^{\circ}$, respectively. 

In the bottom panel of Fig.~\ref{fig:no_corr} the dust ring reported in \citet{Tahani2022} can be seen in the proximity of the OMC.

\section{Cosmic-ray electron energy spectrum}\label{app:CRe}

In this Appendix we summarize and detail the parametric model for the energy flux of CRe -- $j_e(E)$ -- from \citet{Padovani2018} that was used in this work to propose a possible mechanism for the $\beta$ distribution of synchrotron emissivity in the Orion-Taurus ridge (see Sect.~\ref{sec:spec}).
\subsection{Explaining the $\beta$ distribution}\label{ssec:beta}
The synchrotron $T_{\rm b}(\nu)$ is proportional to the integral along the LOS of the sum of the specific emissivities polarized along and across $\vec{B_{\perp}}$, namely $\epsilon_{\nu,\parallel}$ and $\epsilon_{\nu,\perp}$ \citep{Ginzburg1965,Padovani2021,Bracco2022}. These two quantities depend upon the CRe energy spectrum, given by $j_e(E)$, and the value of $B_{\perp}(\vec{r})$ at position $\vec{r}$. Changing $j_e(E)$ and $B_{\perp}(\vec{r})$, one can control the frequency dependence of $T_{\rm b}(\nu)$, thus the value of $\beta$.

\begin{figure}[!h]
\begin{center}
\resizebox{1\hsize}{!}{\includegraphics{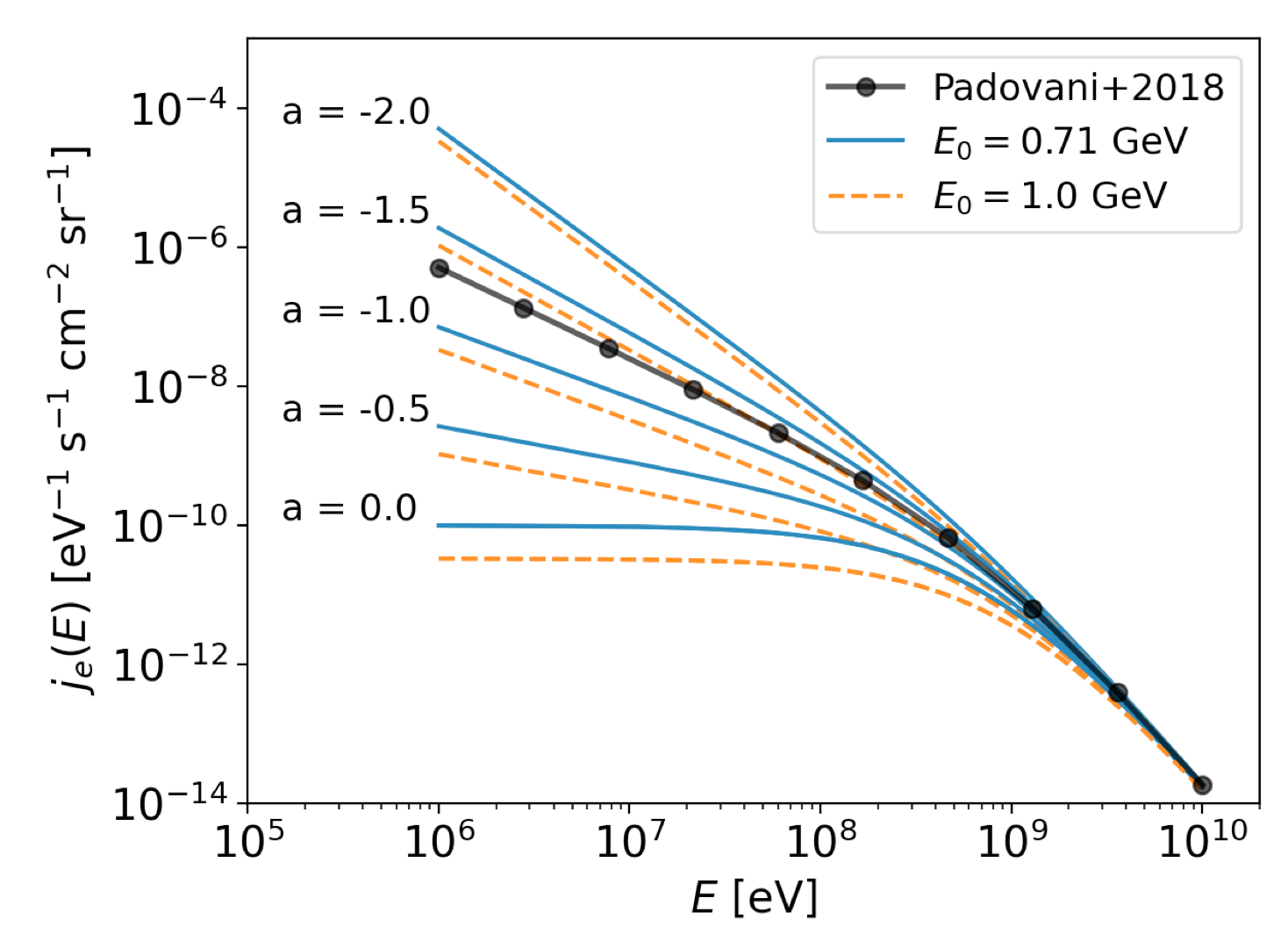}}
\caption{Parametric models for the CRe energy spectrum from \citet{Padovani2018}. See Eq.~\ref{eq:je} for reference.}
\label{fig:je}
\end{center}
\end{figure}

Using state-of-the-art magneto-hydrodynamical numerical simulations of the multiphase ISM, \citet{Padovani2021} thoroughly explored the properties of $\beta$ as a function of $B_{\perp}(\vec{r})$ and $j_e(E)$. They concluded that the shape of $j_e(E)$ has a particularly strong impact on actual observables of synchrotron emissivity at low frequency ($\lesssim$ 10 GHz). In their work the authors mostly relied on the local $j_e(E)$ deduced in \citet{Padovani2018} from real data of interstellar CRe with $E$ between a few hundred MeV \citep[Voyager 1-2,][]{Cummings2016,Stone2019} and hundreds of GeV \citep[Fermi-LAT, Pamela, and AMS,][respectively]{Ackermann2010,Adriani2011,Aguilar2014}. 

\begin{figure}[!h]
\begin{center}
\resizebox{1\hsize}{!}{\includegraphics{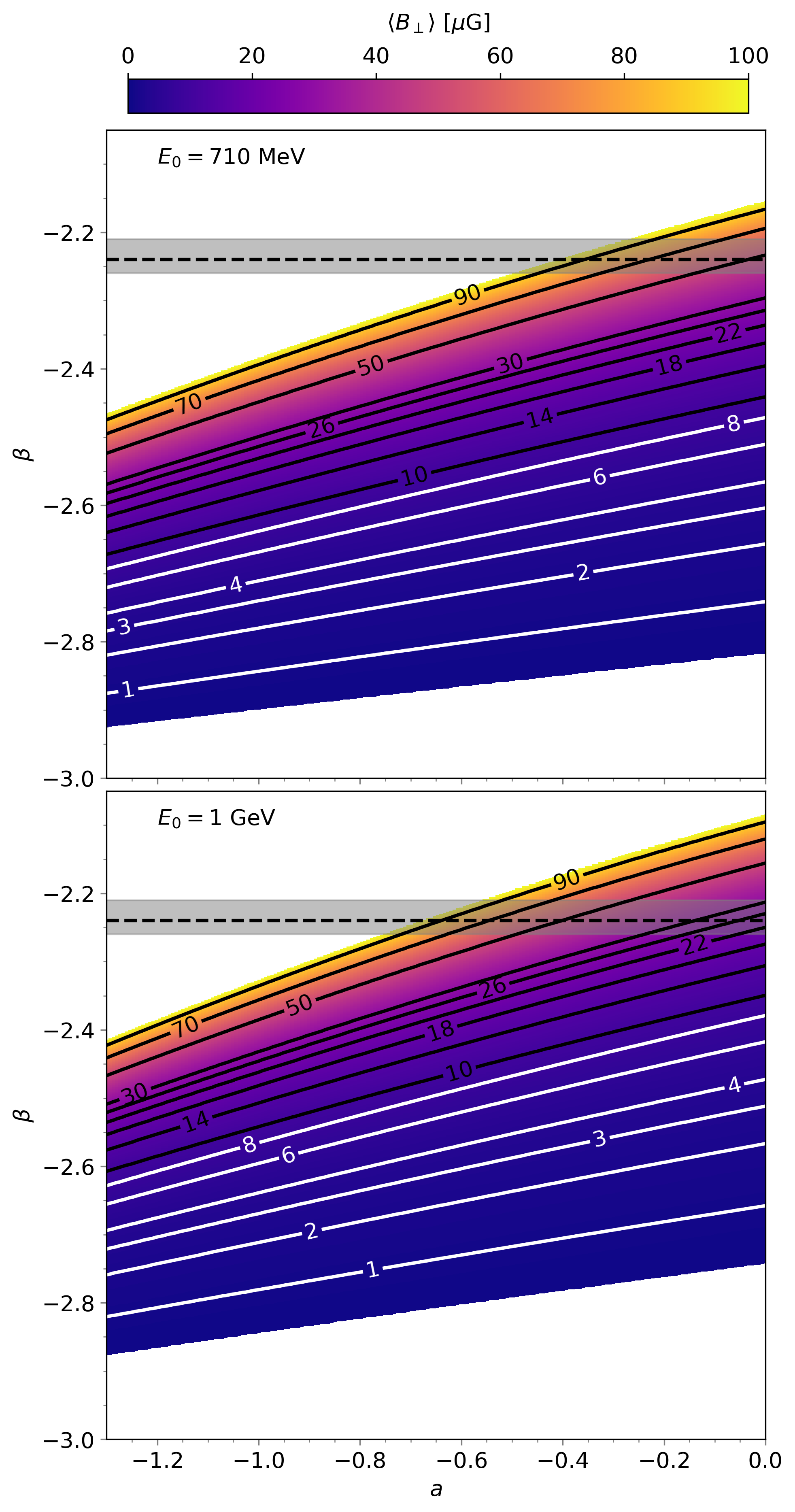}}%{beta_a_v1.png}}
\caption{2D correlation plots between $\beta$ and $a$ as a function of the averaged $B_{\perp}$ along the sight line, $\langle B_{\perp} \rangle$ (in colours and contours). In the top panel $E_0 = 710$ MeV; in the bottom panel, $E_0 = 1$ GeV. The grey-shaded areas and the dahed-black lines are $\beta^{\rm c}_{\rm m} = -2.24^{+0.03}_{-0.02}$ (see Sect.~\ref{sec:spec}).}
\label{fig:beta_a}
\end{center}
\end{figure}

We noticed that by applying the same $j_e(E)$ as in \citet{Padovani2018} -- hereafter, $j_{e0}$ -- we could not reproduce the $\beta$ distribution of the Orion-Taurus ridge measured below 100 MHz with the LWA. In \citet{Padovani2018}, $j_{e0}$ is parameterized as follows:
\begin{equation}\label{eq:je}
    j_{e0}(E) = C\frac{E^{a}}{(E+E_0)^{(3.2+a)}}, 
\end{equation}
where $C = 2.1\times 10^{18}$\,eV$^{-1}$\,s$^{-1}$\,cm$^{-2}$ sr$^{-1}$, $a$\,$=$\,$-1.3$, and $E_0$\,$=$\,$710$\,MeV. 

In this work, we explored the possibility of changing $a$ and $E_0$ in order to obtain values of $\beta$ consistent with $\beta^{\rm c}_{\rm m}$ (see Sect.~\ref{sec:spec}). In Fig.~\ref{fig:je}, we illustrate the $j_e(E)$ variation as a function of $a$ and $E_0$ compared to $j_{e0}$ (see black circles). While $a$ impacts the slope of $j_e(E)$ at low energy, $E_0$ may also slightly affect the high-energy end of $j_e(E)$. Using such analytical models, as shown in Fig.~\ref{fig:beta_a}, we explored a wide range of parameters. We could reproduce the flat $\beta$ observed between 50 and 80 MHz (see the grey horizontal band in Fig.~\ref{fig:beta_a}) by only considering values of $a \gtrsim -0.5$ (lower population of CRe below a few hundred MeV compared to $j_{e0}$) and averaged $B_{\perp}$ along the sight line, $\langle B_{\perp} \rangle$, larger than 30 $\mu$G. This conclusion holds for both $E_0 =710$ MeV and  $E_0 =1$ GeV. We note that we could also reach $\beta$ values consistent with $\beta^{\rm c}_{\rm m}$ even for smaller values of $a$ and $\langle B_{\perp} \rangle$ if only $E_0$ was larger. Looking at Fig.~\ref{fig:je}, this would correspond to having $j_{e}$ flatter than $j_{e0}$ at high energy too. As explained by \citet{Morlino2021}, this effect could also be expected in SN remnants at high energies larger than 10 TeV, where a flattening of the CRe energy spectrum is predicted as the result of electrons escaping from the shock precursors. However, high-energy Fermi-LAT observations suggest that the CR-proton energy spectrum of the OES does not show significant deviations from the local interstellar one \cite[i.e.,][]{Joubaud2020}. 
If this constraint holds for CRe at high energy, the CRe energy spectrum at low energies ($<$ GeV) may be different due to, for example, large energy losses caused by the enhanced magnetic field in OES \citep{Morlino2021}.
A more detailed analysis of the high-energy CRe is left to follow-up work since their impact on the low-frequency radio emission is rather negligible \citep[e.g. Fig.~3 in][]{Padovani2021}. 

\begin{figure}[!h]
\begin{center}
\resizebox{1\hsize}{!}{\includegraphics{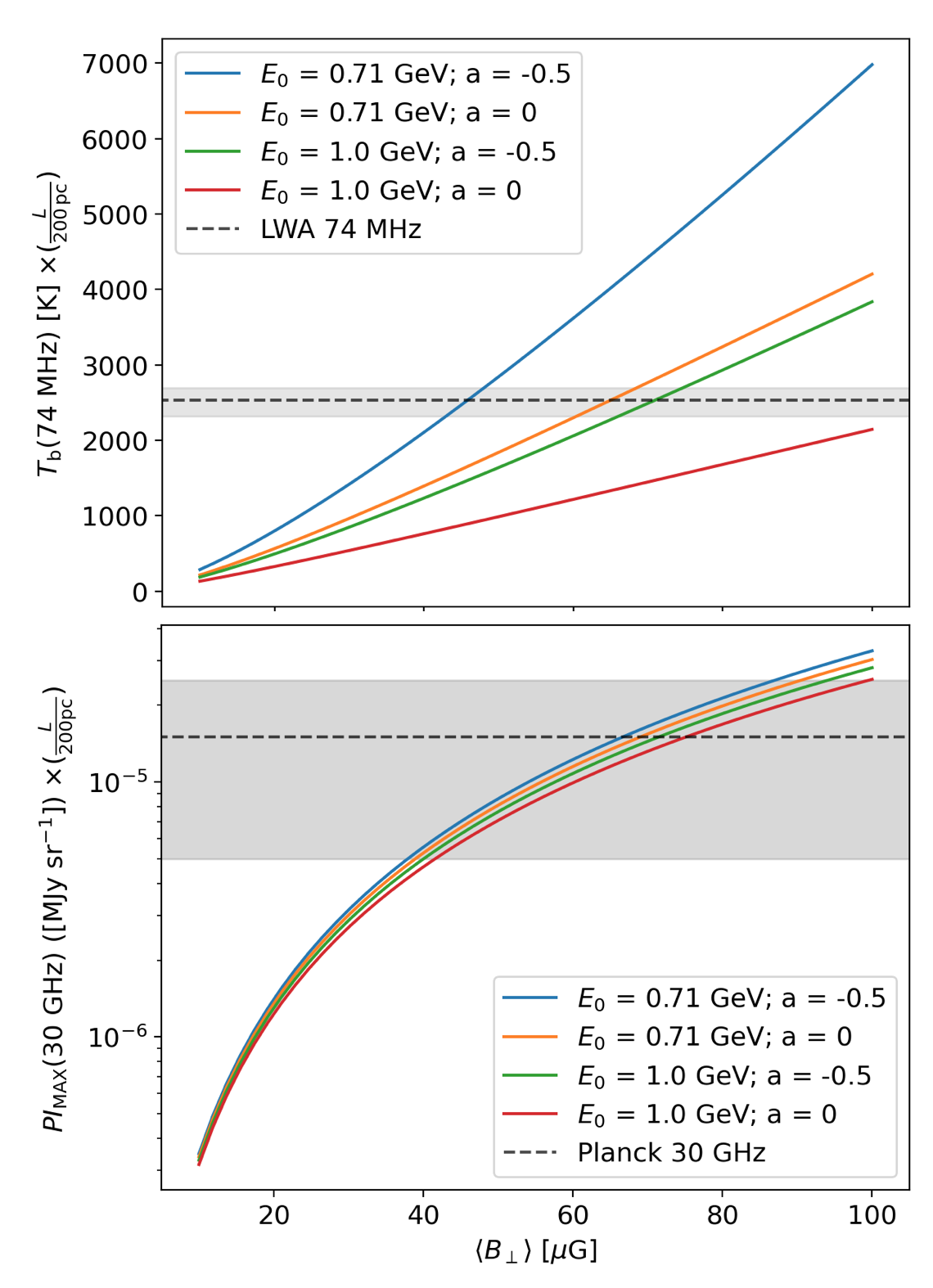}}
\caption{Estimates of $T_{\rm b}$(74 MHz) -- top -- and $PI_{\rm MAX}$(30 GHz) -- bottom -- of the Orion-Taurus ridge from models of \citet{Padovani2018, Padovani2021}. The quantities are derived as a function of $\langle B_{\perp} \rangle$, $a$, and $E_0$. The horizontal black dashed lines, with grey errors, represent the corresponding median values reported in Sect.~\ref{sec:spec}.}
\label{fig:PI}
\end{center}
\end{figure}
\subsection{Synchrotron intensity estimates}\label{ssec:sync}
Our results allowed us to give order-of-magnitude estimates of the synchrotron emission from the Orion-Taurus ridge that we could compare with LWA and {\it Planck} observations (see Sec.~\ref{sec:data}). We refer to the analytical formulation described by Eqs.~(2-7) of \citet{Padovani2021}, which explains in detail how $T_{\rm b}$ and $PI$ can be derived from $\epsilon_{\nu,\perp}$ and $\epsilon_{\nu,\parallel}$ given $B_{\perp}$, $j_{e}$, and the observational frequency $\nu$ \citep[see also,][]{Ginzburg1965}. In short, assuming a homogeneous distribution of CRe and magnetic fields, we made use of the following relations, $T_{\rm b}$\,$\propto$\,$\langle \epsilon_{\nu,\parallel} + \epsilon_{\nu,\perp} \rangle$ and $PI \propto \langle \epsilon_{\nu,\perp} - \epsilon_{\nu,\parallel} \rangle$, where brackets indicate averages along the LOS. For consistency with the 3D dust map, we considered a LOS path-length ($L$) of 200\,pc. The estimates that we obtained represent upper limits to $T_{\rm b}$ and $PI$ at 74\,MHz and 30\,GHz, respectively, as most likely the Orion-Taurus ridge is not uniformly distributed along $L$.

We also note that in the case of $PI$ we provided maximum polarization estimates ($PI_{\rm MAX}$) without including realistic LOS depolarization effects \citep{mamd2008,Gaensler2011}. The tangling of the magnetic field lines along the LOS and within the telescope beam reduces the amount of observed polarization. Depending on the LOS depth and on the turbulent component of the magnetic field, the drop in the observed polarization can range from a fraction to tens of percent for a LOS extent of $L$\,$=$\,$200$\,pc \citep[see Fig.~6 in][]{mamd2008}. In our case, however, the polarization fraction ($PI/I$, where $I$ is the total intensity) of all synchrotron models was the intrinsic theoretical value of 75\% \citep{Rybicki1979}. We neglected the effect of Faraday rotation, which is unimportant at 30\,GHz \citep{Ferriere2020}. 

Figure~\ref{fig:PI} shows our estimates of $T_{\rm b}$(74 MHz) and $PI_{\rm MAX}$(30 GHz) after applying a beam of $2^{\circ}$ FWHM resolution. 
Despite all caveats listed above, we have shown that given magnetic-field strengths of a few tens of $\mu$G ($> 40 \,\mu$G) and the CRe energy properties of the Orion-Taurus ridge derived in Appendix~\ref{ssec:beta}, we could retrieve the right orders of magnitude of both $T_{\rm b}$(74 MHz) and $PI$(30 GHz) from LWA and {\it Planck}, respectively. 

\end{document}